\documentclass[preprint,aps]{revtex4}
\usepackage{graphicx}%
\usepackage{dcolumn}
\usepackage{amsmath}
\usepackage{longtable}
\begin{document}
\title
{ Density functional calculations of atomic structure, charging effect, and 
static dielectric constant of two-dimensional systems based on B-splines } 
\author
{Chung-Yuan Ren$^{a,\dagger}$ and  Yia-Chung Chang$^{b,c}$}
\affiliation
{ $^{a}$ Department of Physics, National Kaohsiung Normal University,
Kaohsiung 824, Taiwan  	\\
$^{b}$ Research Center for Applied Sciences, Academia Sinica, Taipei 115, Taiwan \\
$^{c}$ {Department of Physics, National Cheng-Kung University, Tainan 701, Taiwan}\\
$\dagger$ {\it E-mail address:} cyren@nknu.edu.tw }

\begin{abstract}
We implement a total-energy minimization scheme to allow for relaxation 
of atomic positions in density functional calculations for 
two-dimensional (2D) systems using a mixed basis set. The basis functions  
consist of products of 2D plane waves in the plane of the material 
and localized B-splines along the perpendicular direction. 
By using this mixed basis approach (MBA), we studied the atomic relaxation 
and charge polarization of 2D systems under an applied electric field. 
Compared to the conventional supercell approach (SCA) 
which adopts repeated slabs sandwiched between vacuum regions, 
MBA makes no requirement of compensating background charge 
for treating electrically charged 2D systems due to carrier injection. 
Furthermore, with MBA, the sawtooth potentials for systems under  
the applied field to maintain periodicity as needed in SCA 
is automatically avoided.
From the linear response of charge polarization to the applied  
field, we introduced a simple method to determine 
the out-of-plane dielectric constants of 2D materials without 
the ambiguity of defining their effective thickness. 
Selected 2D systems 
including graphene and transition-metal dichalcogenides are tested.
Our MBA results are consistent with previous SCA calculations when 
both approaches are equally applicable. However, for the charged system 
with high carrier density, we found significant deviation from SCA results 
obtained by imposing artificial charge neutrality condition.	
\end{abstract}
\maketitle
\section{INTRODUCTION}
It is well known that 
the electronic properties of low-dimensional systems are fundamentally
different from those in higher dimensions due to their unusual collective
excitations. Two-dimensional (2D) nano-materials
can be easily fabricated because of the emerging nanotechnology and 
have been attracting much attention in fields of both
theoretical and applied science for material innovation.

First-principles methods based on density functional theory (DFT) with 
pseudopotential (PP) scheme have been
used widely in the electronic structure calculations of solid-state physics and 
quantum chemistry.
In many calculations, three-dimensional (3D) plane waves (PW) are used as
basis functions, which are suitable for bulk systems. 
Because PWs are easily implemented, 
not atom-centered, and can systematically achieve convergence,
they are often employed to expand
the wavefunction even along the non-periodic direction in 2D 
systems by the supercell approach (SCA). 
In this respect, the physical 2D system is treated as a fully
3D periodic system by introducing artificial vacuum space to separate
the repeated slabs along the direction, which should be
considered as nonperiodic. However, SCA 
requires large enough thickness of the vacuum layer 
such that the interactions between the adjacent slabs 
are negligible, and therefore 
increases the number of PW along that direction. 

More seriously, for electrically charged systems 
(or systems with charged defects), 
the long-range tail of 
the Coulomb potential inevitably requires an extremely large separation of the two slabs and 
makes the calculation impractical. This would even cause the 
convergence problem due to the infinitely periodic array of charged defects, 
no matter how large the unit cell was chosen. One simple way to avoid this
issue is the use of charge neutrality condition by adding an additional 
compensating charge to make the whole system electrically neutral. 
Several more sophisticated correction schemes have also been devised to 
remedy the difficulty \cite{TB}-\cite{IB}. 
Another drawback of SCA is the unavoidable discontinuity of the sawtooth 
potential in the system under an external E-field \cite{MV,Beng} or an 
asymmetric slab with a net surface dipole density 
(like ferroelectric BaTiO$_3$) \cite{MV}. Such  
discontinuity still persists with the proposed
dipole correction \cite{Beng} 
and should be placed within the vacuum region, 
where physical quantities of interest are ensured to be negligible.

In previous work \cite{LC}-\cite{RCH}, a mixed basis approach (MBA)
has been introduced for the first-principles calculations
of low-dimensional systems 
by expanding the wavefunction along the periodic direction with
PW but along the finite non-periodic direction with
localized basis functions. MBA uses only one slab that contains the physical 
atom layers and some nearby vacuum space to allocate BS basis functions. 
MBA has several advantages over SCA:
(1) In MBA, each unit cell corresponds to the true unit cell of the  
 real 2D material. MBA 
 retains the layer-like local geometry as in the real physical surfaces.
(2) Instead of using alternating slabs and vacuum regions in SCA,
one can directly calculate the energetics, structure, and dynamics of 
an isolated slab without any correction.
(3) For charged systems, the spurious Coulomb interaction between the excess 
charge,
its images and the compensating background charge in SCA can
be automatically avoided.  
(4) In an external electric field or with surface dipole moments,
MBA needs no inclusion of dipole-corrected slabs. 
(5) The number of the basis could be  
reduced, easing the computational burden for the
diagonalization of the Kohn-Sham Hamiltonian.
MBA, conceptually very simple, is suitable for
investigating low-dimensional systems 
including surfaces, interfaces, and superlattices. 

As demonstrated elsewhere\cite{RHC}-\cite{RJH}, B-splines (BSs) are well 
suited to describe the localized wavefunction. We choose BS as 
the basis in MBA to expand the Kohn-Sham orbitals   
perpendicular to the surface. 
BSs are highly localized piecewise polynomials within
prescribed break points. 
BS has several desired properties:
(1) BSs and their derivatives 
can be evaluated easily and precisely.
(2) BSs possess good flexibility to represent a rapidly varying
wavefunction accurately by adjusting the break points to
have an optimized basis. (3)   
Unlike Gaussian functions and atomic orbitals which are 
an atom-centered basis, BSs are independent of
atomic positions \cite{AG,HGG}, 
so the atomic-structure optimization can be implemented without complexity.

We notice that one B-spline-based finite-element (FE) approach 
was developed to achieve chemical accuracy efficiently \cite{TMG}.
FE avoids transformation into the reciprocal space 
and its numerical efforts are linear with the system size \cite{Geo}. 
Such approach has proven to be efficient for polyatomic
molecules/clusters and allows for more flexible boundary conditions 
to the solution of Poisson equation \cite{WWT}-\cite{BHW}  
(an overview of FE can be found in Refs. \cite{TMG,Geo,book2}).
Therefore, our MBA-BS, with local discretization refinement embedded in FE,
would be beneficial to one dimensional systems (e.g. an infinitely long
graphene nanoribbon)
where the real-space integration involves two non-periodic directions.

In this paper, based on DFT with Vanderbilt's 
ultra-soft pseudopotential (USPP) \cite{DV},
we extend our previous MBA approach \cite{RHC} 
to study the atomic and electronic 
structures of selected 2D systems, particularly for charged systems and systems 
under an applied E-field.
We examine the atomic relaxation, the charging effect due to carrier injection,
and the static dielectric constant within the MBA-BS scheme.
This paper is organized as follows:  In Sec. II, the
computational method is presented. 
In Sec. III, we report the results of practical tests, which  
demonstrate explicitly the MBA capability. The results 
are displayed and discussed. Finally, the summary is given in Sec. IV.
The relevant details of the total energy and force formula 
in terms of BS for 2D systems are shown in the Appendix.

%
\section{METHOD OF CALCULATION}
\subsection{B-splines}
For the sake of completeness, we briefly summarize the BS formalism.
More details can be found in Refs. \cite{RHC,deBoor}.

BSs of order $\kappa$, 
$
\{ B_{i,\kappa}(z)\}_{i= 1, \ldots},
$
are determined by a sequence of nondecreasing numbers $\{\tau \}$
which is referred to as a knot sequence.
$
\{B_{i,\kappa}(z) \}
$
is  a set of 
locally positive polynomials of degree $\kappa -1$ with
compact support $\tau_i \le z\le \tau_{i+\kappa}$ and 
vanish everywhere outside those subintervals. 

BS is generated by the recursive relation :
\begin{equation}
B_{i,\kappa}(z)=\frac{z-\tau_i}{\tau_{i+\kappa-1}-\tau_i}B_{i,\kappa-1}(z)+\frac{\tau_{i+\kappa}-z}{\tau_{i+\kappa}
-\tau_{i+1}}B_{i+1,\kappa-1}(z),
\end{equation}
with
\begin{equation}
B_{i,1}(z)= \left \{ \begin{array}{ll}
                1,      & \tau_i \leq z < \tau_{i+1}  \\
                0,      & {\rm otherwise \ .}
                \end{array}
                \right.
\end{equation}
Its first derivative is given by
\begin{equation}
\frac{d}{dz}B_{i,\kappa}(z)=\frac{\kappa-1}{\tau_{i+\kappa-1}-\tau_i}B_{i,\kappa-1}(z)-\frac{\kappa-1}
{\tau_{i+\kappa}-\tau_{i+1}}B_{i+1,\kappa-1}(z).
\end{equation}
Therefore, the derivative of BSs of order $\kappa$ is simply a 
linear combination of BS of order $\kappa-1$, which is
also a simple polynomial and is continuous across the knot sequence.
The flexibility of BS to accurately represent localized 
functions was demonstrated in Ref. \cite{RHC}.
In practical calculations, we use $\kappa=4$, that is, 
BSs are cubic polynomials. 

\subsection{Hamiltonian and total energy} \label{section}
Atomic units, $m=\hbar=e=1$, are used throughout this paper. 

With BS for the non-periodic $z$ 
direction and 2D PW for the periodic $xy$ plane,
the mixed basis used to expand the wavefuction
is defined as
\begin{equation}
< {\bf r } | { \bf k +  g} ; j, \kappa > \
=
\frac{1}{\sqrt{A}}\
e^{i( {\bf k +  g} ) \cdot { \bf \rho } }
\ B_{j,\kappa}(z),
\end{equation}
where $ {\bf g} $
denotes 2D reciprocal lattice vector and   
$ {\bf k} $ is Bloch wave vector.
$A$ is the surface area of the system.
Therefore, the charge density can be written in the form  
\begin{equation}
n({\bf r})  = \sum_{{\bf g}}  
\ n({\bf g},z)
\  e^{ i {\bf g} \cdot {\bf \rho} } \ ,
\end{equation}
where the sum runs over up to an appropriate energy cutoff.

In USPP scheme \cite{DV,LPCLV}, the wave function $\phi_i$
satisfies a secular equation of the form
\begin{equation}
H|\phi_i>=\epsilon_i S |\phi_i>, 
\end {equation}
subject to a generalized orthonormality condition
\begin{equation}
<\phi_i|S|\phi_j>=\delta_{ij}\ .
\end{equation}
Here,
\begin{equation}
H=-\frac{1}{2}{\nabla }^2 +V_{\text{eff}}
+\sum_{Inm}D^I_{nm} |\beta^I_n> < \beta^I_m|.  \label {eq1}
\end {equation}
The screened effective local potential 
$V_{\text{eff}}$
includes the local potential part of USPP, Hartree potential, and 
exchange-correlation potential, 
\begin{equation}
V_{\text{eff}}=V_{loc}+V_H+v_{xc}.	\label{veff}
\end{equation}
The last term of the right hand side in Eq. (\ref{eq1}) 
is the non-local potential part $V_{NL}$ of USPP
and  
$\beta^I_n$ is the n$^{\text{th}}$ projector function, centered on site $I$.

As for the overlap operator $S$, it is given by
\begin{equation}
S=1+\sum_{Inm}q^I_{nm} |\beta^I_n> < \beta^I_m|, 
\end {equation}
where $q^I_{nm}=
\int d{\bf r} Q^I_{nm}({\bf r})$. The augmentation functions 
$ Q^I_{nm}({\bf r})$, also centered on site $I$, are strictly localized in 
core regions. Note that $D^I_{nm}$ in Eq. (\ref{eq1}) should be 
determined self-consistently via 
\begin{equation}
D^I_{nm}=D^0_{nm}+
\int d{\bf r} Q^I_{nm}({\bf r}) V_{\text{eff}}({\bf r}),  
\end {equation}
where the strength $D^0_{nm}$ is provided by USPP and 
differs for different ion species.
The charge density from the wave function is augmented inside the
core region,
\begin{equation}
n({\bf r}) =\sum_i |\phi_i({\bf r})|^2 +
\sum_{Inmi} Q^I_{nm}({\bf r})
< \phi_i | \beta^I_n> < \beta^I_m| \phi_i>. 	
\end{equation}
%
%
The total energy $E_{tot}$
is given by
%
%
\begin{equation}
E_{tot}=\sum_{i}w_i<\phi_i|H|\phi_i>
-\frac{1}{2}\int d {\bf r} n({\bf r}) V_{H}({\bf r})
- \int d{\bf r} n({\bf r}) v_{xc}({\bf r})
+E_{xc}
+E_{ii}, \label{etot}
\end{equation}
where the sum runs over the occupied states with appropriate weight $w_i$.
Here, $E_{xc}= \int d{\bf r} (n+n_c){\varepsilon}_{xc}[n+n_c]$ is the 
exchange-correlation energy with 
pseudized core charge density $n_c$ 
if nonlinear core correction (NLCC) \cite{LFC} is taken into account. 
$E_{ii}$ denotes the ion-ion repulsive energy. 
When an external electric field ${\bf E}= {\mathcal{E}}_0{\bf \hat{e}_z}$ 
perpendicular to the surface  is applied,  
$E_{tot}$ in Eq. (\ref{etot}) is added with 
\begin{equation}
E_E=\int d {\bf r} n({\bf r})V_E(z)-{\mathcal{E}}_0\sum_I Z_I z_I. 
\label{eq23}
\end{equation}
where 
$V_E(z)=-{\mathcal{E}}_0 z$. $Z_I$ and $z_I$ are 
the ionic charge and $z$-coordinate of ion $I$, respectively.
%
\subsection{Force} \label{section}
The forces are defined as the total derivative of the total energy with respect 
to ionic positions ${\bf R_I}$, 
\begin{equation}
{ \bf F}=-\frac{dE_{tot}}{d {\bf R_I}}.
\end{equation}
To demonstrate the flexibility of BSs, we only focus on the $z$-component.  

Using Hellmann-Feynman theory, the force within USPP scheme is 
\begin{equation}
F_z=-\sum_{i}w_i<\phi_i|\frac{\partial (H-\epsilon_i S)}
{\partial z_I}|\phi_i>
-\frac{\partial E_{xc}}{\partial z_I} 
-\frac{\partial E_{ii}}{\partial z_I}. 	\label{eqf}
\end{equation}
Note that we need not to calculate the change of $V_H$ or $v_{xc}$ 
in $H$ due to the change 
of the soft or the augmented charge because, to the first-order, the change of
the sum of Kohn-Sham eigenvalues will cancel out the change of these potential
contributions \cite{KJ}. It turns out that there are several contributions to 
the total force. The first term is  
\begin{equation}
F^{loc}_{Iz}= -\int d {\bf r} n({\bf r}) 
\frac{\partial V_{loc}( {\bf r})}{\partial z_I}.  \label{eq10}
\end{equation}
The second term arises from $D_{mn}$ due to the change of the augmentation
charge when the ion is moving, 
\begin{equation}
F^{ln,1}_{Iz}=-\sum_{nmi}
\left [ \int d{\bf r} \frac{\partial Q^I_{nm}({\bf r})}{\partial z_I}  V_{\text{eff}}({\bf r}) \right ] 
\ \omega_i
< \phi_i | {\beta}^I_n> < \beta^I_m| \phi_i>. \	\label{eq11} \\
\end{equation}
The third one is due to the change of the projector and is given by
\begin{equation}
F^{ln,2}_{Iz}=-\sum_{nmi} \omega_i (D^I_{nm}-q^I_{nm}\epsilon_i)
< \phi_i | \frac{ \partial |\beta^I_n> < \beta^I_m|}{\partial z_I}| 
\phi_i> \ \label{eq12}	\\
\end{equation}
In the $-{\partial E_{xc}}/{\partial z_I}$ term in 
Eq. (\ref{eqf}), the force due to the change of
the frozen pseudized core charge $n_c$ is  
\begin{equation}
F_{Iz}^{nlcc}=- \int d {\bf r} v_{xc}({\bf r}) 
\frac{\partial n_c^{I}( {\bf r})}{\partial z_I}.  
\end{equation}
The ion-ion force can be treated by the Ewald sum \cite{Kax,Koh}. 
The detailed mathematical derivations of the relevant force components, 
as well as $H(S)|\phi_i>$ based on MBA-BS for 2D systems 
will be given in the Appendix.  

We used the conjugate-gradient algorithm implemented previously 
\cite{RCH1} for the eigenvector/eigenvalue searching. 
The geometry optimization was performed with th 
Broyden-Fletcher-Goldfarb-Shanno (BFGS) algorithm \cite{bfgs}
to find the total energy 
minimum.   
\subsection{No charge neutrality condition}	\label{ncn}
The electrostatic potential $V_c$ in 2D momentum representation 
is
\begin{equation}
V_c({\bf g}\neq 0,z)= \int d z' n_t({\bf g},z')
\frac{2 \pi}{|{\bf g|}} \
e^{- |{\bf g}| |z-z'|}. 	\label{eq2d}
\end{equation}
Here, $n_t$ consists of electron and ion charge density. 

For $|{\bf g}| \rightarrow 0$, 
\begin{eqnarray}
V_c({\bf 0},z) &=& \int d z' n_t({\bf 0},z')
\frac{2 \pi}{|{\bf g|}} \
(1-|{\bf g}| |z-z'|)	\nonumber \\
&=& \frac{2 \pi}{|{\bf g|}} \int d z' n_t({\bf 0},z')-
2 \pi \int d z' n_t({\bf 0},z')|z-z'|	\label{g00}
\end{eqnarray}
The first term of the right hand side in the above equation 
could be safely dropped if the system 
is charge neutral. For the situation of nonzero net charge,
the $V_c({\bf 0},z)$ component is rewritten as 
\begin{eqnarray}
V_c(\boldsymbol {0}, z)	\nonumber 
&=&\iiint \ dz' 
\frac{d^2 \boldsymbol{\rho'}}
{ \sqrt{\Delta {z}^2+{\rho'}^2 }} 
\ n_t({\bf 0},z')  	\nonumber \\ 
&=&\iint_0^R \ dz' \frac{2 \pi \rho' d {\rho'}} 
{\sqrt{\Delta {z}^2+{\rho'}^2 }} 
\ n_t({\bf 0},z'), 
\end{eqnarray}
where 
we assume $R$, the 'radius' of the system along the $xy$ plane be arbitrarily
large but finite, and $\Delta z =z-z'$. Since 
$R$ is much larger than the dimensional size in the $z$ direction, i.e., 
$ R \gg \Delta \boldsymbol {z}$, then   
\begin{eqnarray}
 \int_0^R \frac{\rho' d \rho'} 
{ \sqrt{\Delta {z}^2+{\rho'}^2 }}&=&
\sqrt{R^2+\Delta{z}^2} -|\Delta{z}| 	\nonumber \\
&\sim& R- |\Delta{z}|.
\end{eqnarray}
So,
\begin{equation}
V_c({\bf 0},z)= 
2 \pi R \int d z' n_t({\bf 0},z')-
2 \pi \int d z' n_t({\bf 0},z')|z-z'|.	\label{g01}
\end{equation}
In the case of charge neutrality,
the first term on the right hand side of Eq. (\ref{g01}) vanishes and we
retain Eq. (\ref{g00}). For charged systems with net planar charge density
$ \int d z' n_t({\bf 0},z') \neq 0$,
such term would be huge. But clearly it is a constant that is independent upon
$z$, and only causes a shift to the total energy. This kind of constant is
irrelevant to the band structure calculation. Therefore,
we could, just like the charge-neutral case, omit the
first term  without further corrections.
What we need to care about is 
only the second term on the right hand side of Eq. (\ref{g00})
for both neutral and charged cases within MBA.
\newpage
\section{APPLICATIONS OF PRESENT METHOD} \label{cm}
\subsection{Preliminary numerical test}  
First of all, we take 2H MoS$_2$ monolayer  
to test the MBA performance. In the 2H phase, Mo occupies at
the Wykoff 1c site (0, 0, 0) and S occupies at the 2s sites 
$(1/3, 2/3,\pm u)$. 
We perform total energy calculations 
to find the relaxed internal coordinate $u$. 
The in-plane lattice constant $a_0$ was set to 3.16 \AA.  
The calculations were done with 
25 BSs that are distributed over a range of 4.0 $a_0$ and 
the energy cutoff of 20 Ry for 2D PW.
A Monkhorst-Pack $7 \times 7$ mesh 
including $\Gamma$ point was taken to sample the 2D 
irreducible Brillouin zone (IBZ). 
We used Mo and S USPPs \cite{DV,GBRV}
which were generated from the Vanderbilt's code \cite{vancode}.    
The generalized gradient approximation with 
Perdew-Burke-Ernzerhof exchange-correlation functional \cite{PBE} is adopted. 
The potential is determined self-consistently until its
change is less than $10^{-7}$ Ry. 

\begin{figure}[h]
        \caption{ } \label{fig3}
\includegraphics{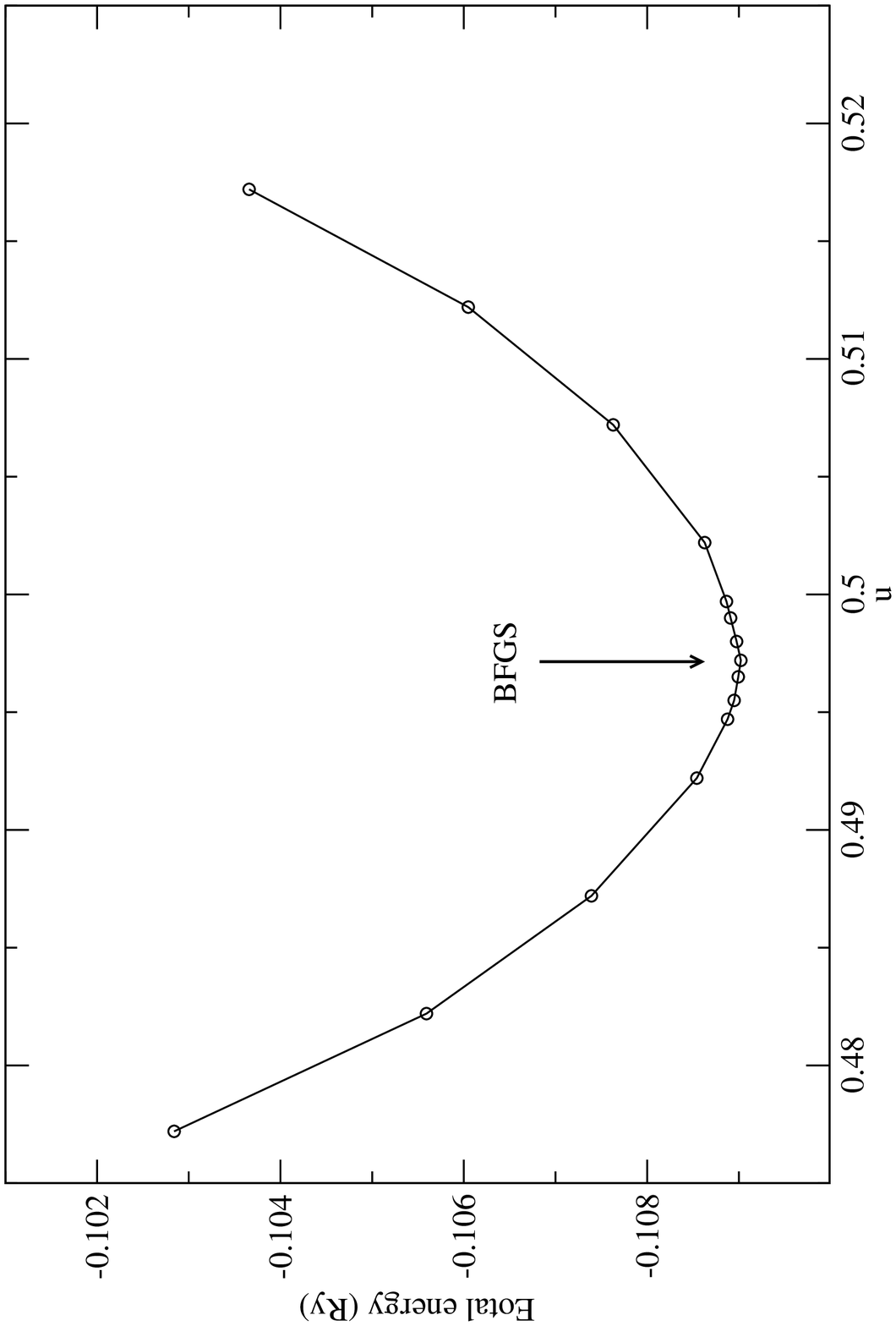}
\end{figure}

Figure \ref{fig3} shows $E_{tot}$ as a function of $u$, 
with the change in steps of $\Delta u \sim0.001 $ near the minimum.  
The energy in this figure is given  relative to some reference energy.
The energy minimum occurs around $u=0.497$, which is very close to
the value of 0.4972 by the BFGS algorithm. 
The residual force along $z$ direction is found to be less 
than 0.01 mRy/a.u. 

We also summarize in Table \ref{tab0} the relevant information for the total 
number of basis functions used by both MBA and SCA. 
It is worth mentioning that the range of vacuum space
in MBA are mainly determined by the wavefunction, which quickly decays outside 
the surface (at least for electrically neutral and positively charged systems). 
On the other hand, the vacuum layers in SCA depend upon the Coulomb potential 
which could exhibit a long-range tail in the vacuum space. 
The reduction in the number of basis by MBA 
will ease the computational efforts for 
the Kohn-Sham eigenvalue searching, which is the dominant  
cost in DFT calculations. We found that the execution time  
for the self-consistent 
loop by the present code is 5.3 seconds per iteration per processor, 
compared to the 1.9 seconds by VASP 
which is well developed and optimized with 3D plane waves. Needless to say,
the comparison for the running time 
will depend on the size of vacuum space set in VASP, 
the real-space grids for FFT, the quality of pseudo-potentials 
(like the number of projectors) and the algorithms used. 
For example, the time is found to be 
2.6 seconds per iteration per processor with an increasing vacuum layer of 
15 \AA~ in VASP. 

In this work, we do not intend to compete in speed with 
the planewave-based codes. 
The advantages of MBA mentioned in the Introduction will come at the price 
of extra real-space integration along the perpendicular direction. 
Maybe other methodologies, e.g., the 3D BS-FE approach \cite{TMG} 
could be studied in the future.
The data presented here are just to give an idea of how our MBA works in speed  
with respect to VASP. 
In any case, the real-space integration is computationally moderate 
because of the characteristics of the smoothness of USPP. In addition, relevant
quantities with the integration can be precalculated to speed
up the calculation in the self-consistent iteration loop. 
\subsection{Transition Metal Dichalcogenides}  
Now, we systematically performed structural optimization for   
transition metal dichalcogenides (TMDC) MX$_2$, 
which have attracted much attention recently \cite{Jiang}.  
TMDC are characterized by their layered structures. 
Here, we study the electronic and structural properties of two    
monolayer MX$_2$ families:
MoS$_2$ family with M=Mo, W and X=S, Se, and  
ZrS$_2$ family with M=Zr, Hf and X=S, Se. 
We focus on the 2H phase only and 
carry out the optimization of $u$   
with experimental lattice constants \cite{SHM}-\cite{HS}.     
The calculation condition is similar to that in 
the above preliminary test. For the ZrS$_2$ family,  
the BS number is slightly increased to 29 to 
account for the larger size.  
For comparison, we also performed calculations by SCA implemented in the 
VASP code with projector-augmented-wave (PAW) potentials 
\cite{KJ,KF}.  
A typical vacuum space of 10-15 \AA~ required in VASP was 
used in the calculation.

Table \ref{tab1} 
summarizes the optimized vertical M-X distance $d_z$ ($ua_0$) and 
band gap $E_g$ of these eight compounds, along with those obtained by 
VASP. Overall, we found an good agreement between the present 
$d_z$ results and those by VASP. 
We also examine with another two 
different BS sets and $d_z$ is almost unaltered. 

\begin{figure}[h]
        \caption{ } \label{mos2}
\includegraphics{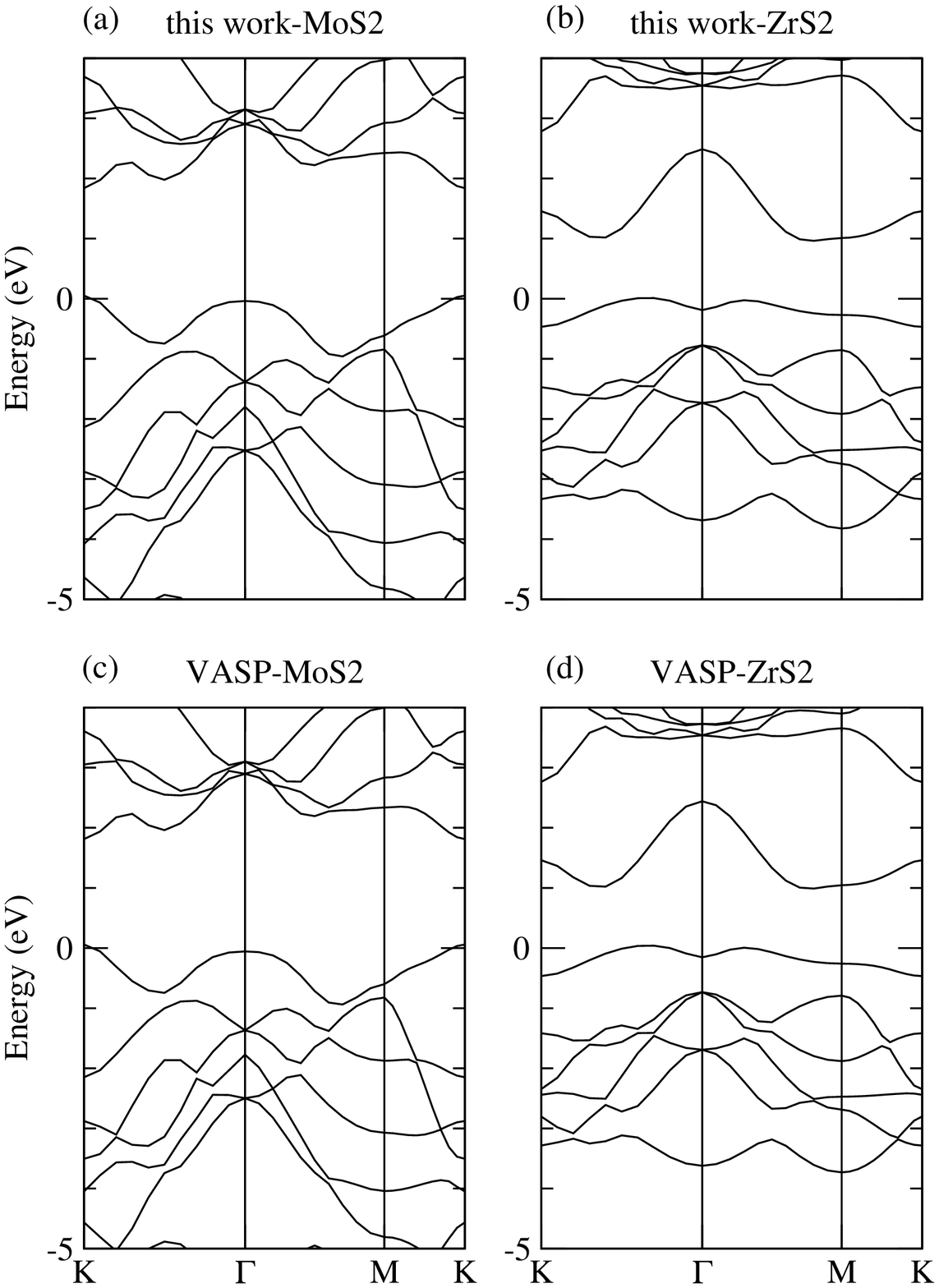}
\end{figure}

As for the electronic structure, we display the band structures of 
MoS$_2$ and ZrS$_2$ in Figs. \ref{mos2}(a) and (b) for the two families. 
Figures \ref{mos2}(c) and (d) show the corresponding VASP counterparts.
Clearly, MoS$_2$ has a direct band gap with both 
valence band maximum and conduction band minimum 
falling at K ({\bf k}=(1/3,1/3)) whereas ZrS$_2$ has an indirect band gap, 
which agree well between MBA and SCA.      
A detailed analysis shows that, for ZrS$_2$, 
valence band maximum occurs within segment K-$\Gamma$, 
and conduction band minimum falls within segment $\Gamma$-M $({\bf k}=(1/2,0))$, 
It can be seen from Table I that $E_g$ obtained 
by MBA is in a quantitative agreement with the VASP result.  
The only slight discrepancy which appears in the HfS$_2$
and HfSe$_2$ cases is attributed to the quality of Hf pseudopotentials used. 
Actually, we have done the calculation with two different Hf USPPs (and PAWs) 
and found that while 
the optimized $d_z$ is almost unaffected, $E_g$ differs significantly
($\sim$0.15 eV), reflecting the sensitivity of the electronic structure to Hf
PP quality.

\subsection{Charged graphene}
Next, we apply the present method to electrically charged systems 
which are very challenging for SCA because of the spurious long-range 
Coulomb interaction between the excess charge and its periodic images 
due to the periodic boundary condition \cite{OS}.  
Naturally, it should remove such boundary condition to study the 
surfaces that are charged up or have dipole moments.
Charged 2D systems can be achieved experimentally via carrier injection in a 
field-effect-transistor setup \cite{SBCSFKDI}.

For simplicity, we use the graphene sheet as a test example.  
All carbon atoms were kept at ideal positions with 
$a_0$ = 2.46 \AA.  
Here, one of every eight valence electrons in the unit cell was removed, 
that is, 
the nominal ionicity of C in this artificial positively-charged system is +0.5. 
13 BSs distributed over a range of 3.25 $a_0$
are used and the energy cutoff of the 2D PW is 30 Ry.
A dense Monkhorst-Pack $21 \times 21$ mesh 
was used to sample the surface IBZ of this metallic system.

\begin{figure}[h]
        \caption{ } \label{charge}
\includegraphics{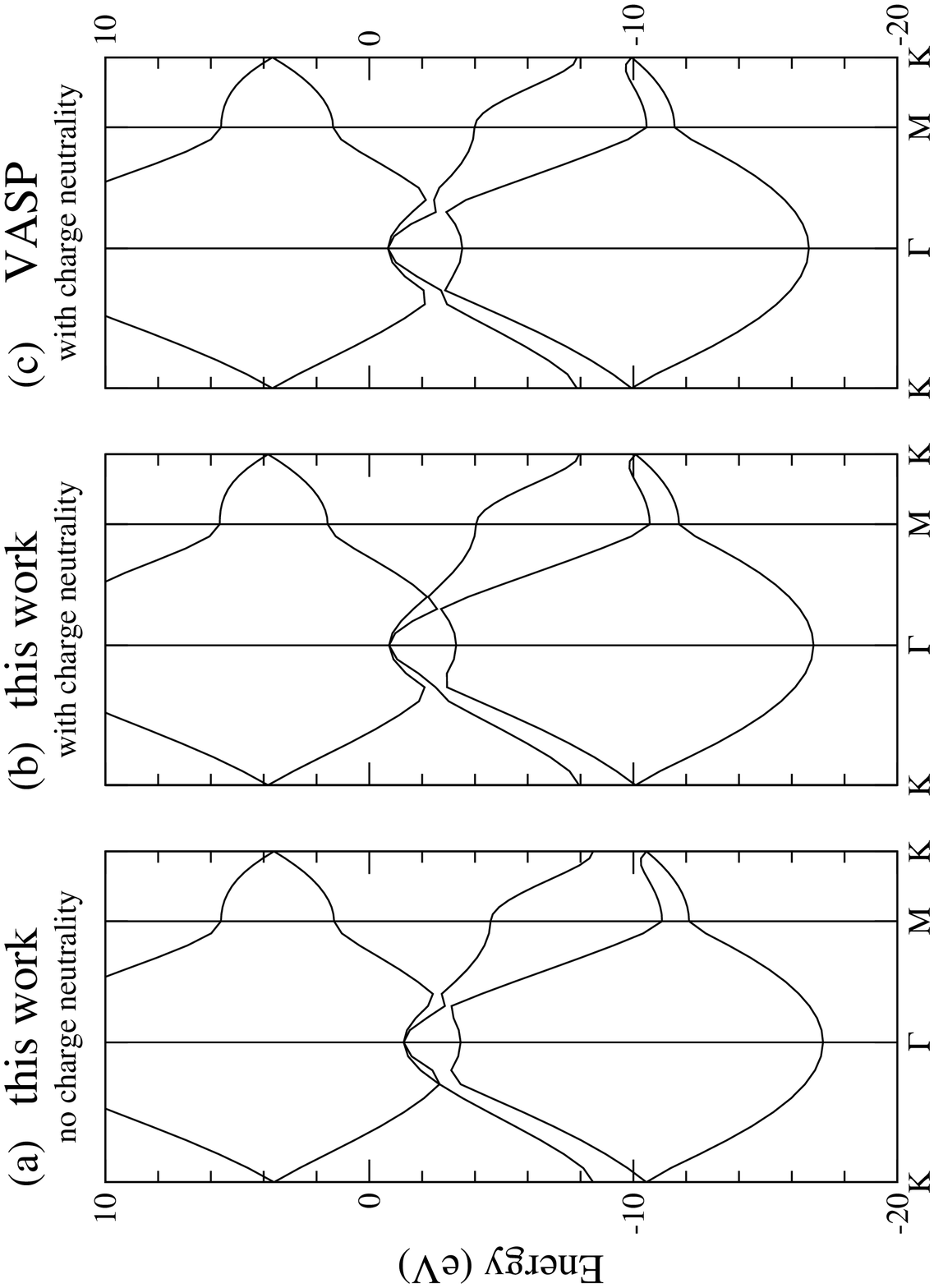}
\end{figure}

The calculated band structure is shown in Fig. \ref{charge}(a). 
For comparison, we also do the calculation 
imposed by the charge neutrality condition with a 
compensating charge. The corresponding result is presented in  
Fig. \ref{charge}(b) and the VASP counterpart in Fig. \ref{charge}(c). 
Clearly, with charge neutrality condition, MBA yields similar 
band structures with the VASP. However, 
the results with and without charge neutrality condition
are significantly different \cite{note1}, particularly those near Fermi level 
around $\Gamma$ point. 

In SCA, the charged defects are unfortunately subjected to 
the spurious image interaction, and no feasible size in practice would be 
sufficient to render this long-ranged electrostatic interaction negligible.  
A cheap way to 
avoid divergence of the electrostatic energy is to impose an additional 
compensating charge into the system.
In some experiments, the surrounding medium (like metal or solution) 
around these excess charges would change
its electron spatial distribution to perfectly screen the defects so that the 
the above imposition was justified \cite{OS}. However, 
in some situations, the 'reference' electrode is put sufficiently 
far away from the system interested, for example, the charged surface of 
the Van de Graaff sphere or the rubbed plastic plate with a net static 
electricity. The often imposed 
charge neutrality condition would not be valid.   

It's true that the Coulomb potential would diverge even along 
an infinitely ideal 
charged plane. But, actually in reality all physics systems are finite, 
e.g., the rubbed plastic plate. How can we investigate such a system?
Clearly, with the size of planar charged systems being arbitrarily 
large but {\it finite}, the first term in Eq. (\ref{g00}) or (25) 
can be safely dropped out, as explained in Section \ref{ncn}, and 
our method can be used to mimic this kind of systems except for the edge 
effect. All we have to do is to evaluate the second term of Eq. (\ref{g00}) no 
matter the system has an excess charge or not.  
No further corrections are needed in MBA since only one single isolated
charged slab rather than an array of the replicated ones is under 
consideration.   

The key point of the present method is the utility of Eq. (\ref{eq2d}) 
for Coulomb potential for 2D systems instead of 
the usual expression of $\sim n_t({\bf G})/|{\bf G}|^2$ for 3D systems
(${\bf G}$ denotes a 3D reciprocal lattice vector). 
It may be argued that the large constant in the first term of Eq. (25) 
can also be dropped in the 3D plane-wave code without affecting the 
relative band energies. However, to evaluate the second term in Eq. (25), 
some remedy will be required to subtract the artificial contributions 
generated from periodically repeated charged sheets included in the 3D 
supercell method.

\begin{figure}[h]
        \caption{ } \label{converge}
\includegraphics{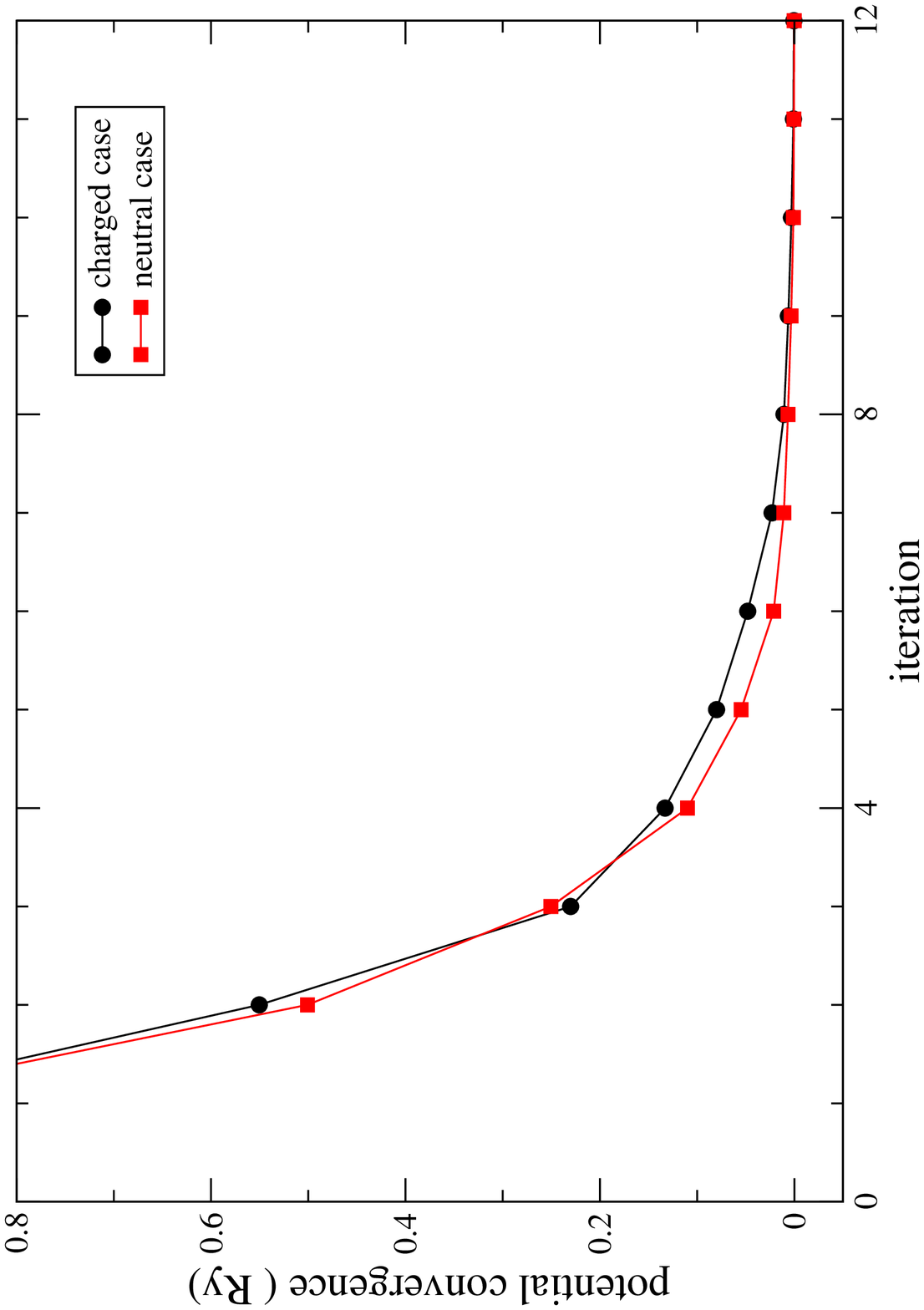}
\end{figure}

Note that the convergence rate of the calculation 
is stable and as fast as for the neutral case, 
as shown in Fig. \ref{converge}.   
To our knowledge, it seems that the existing packages based on SCA 
could not deal with such situation where the charge 
neutrality condition was unsatisfied. 
MBA provides an alternative way to study 
both neutral and charged 2D systems with no complications. 
\subsection{Bilayer graphene under an external E field} \label{2ghE}
\begin{figure}[h]
        \caption{ } \label{ab_E}
\includegraphics{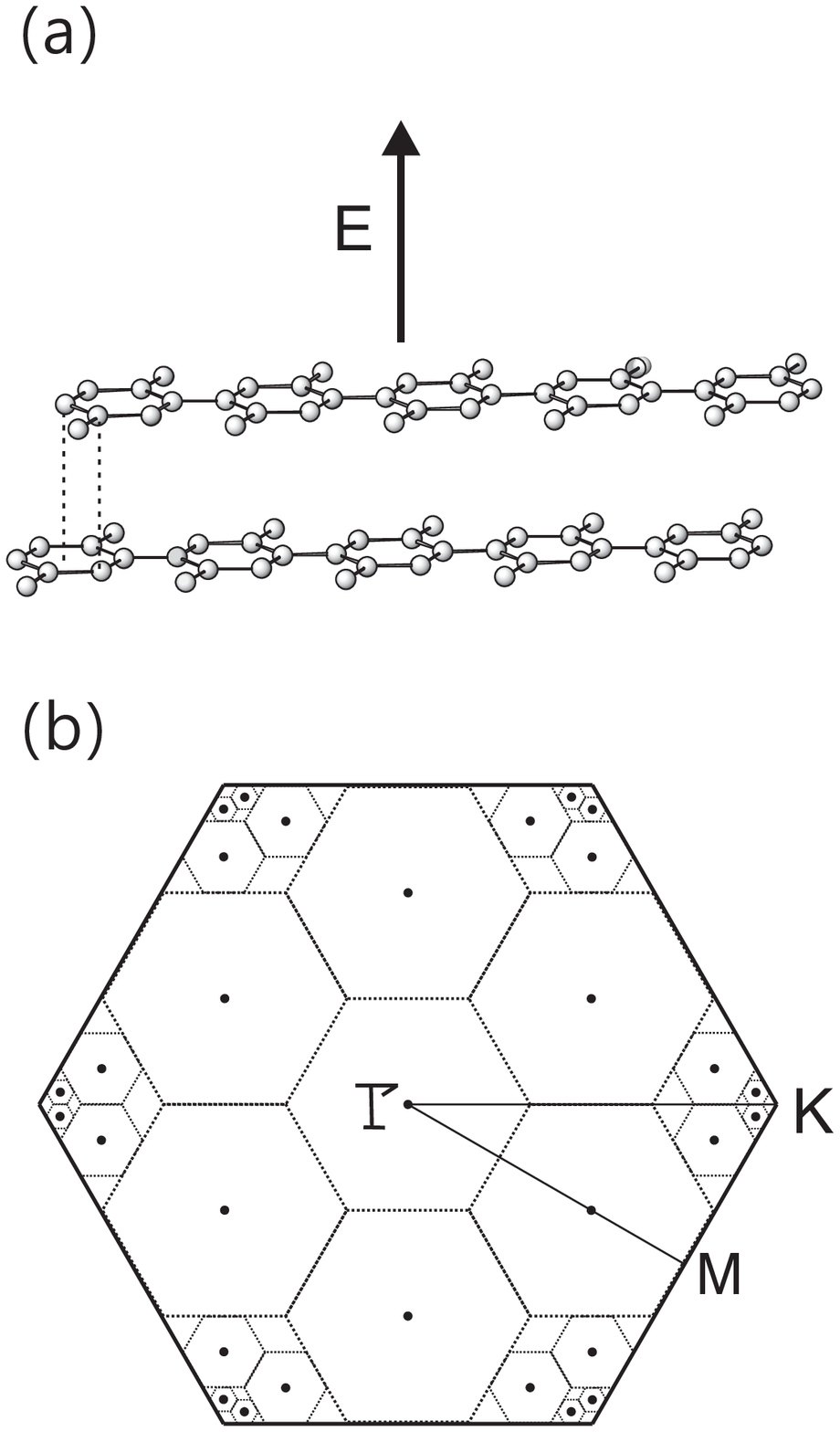}
\end{figure}

The third example is the system of bilayer graphene under 
an external E-field, as shown in Fig. \ref{ab_E}(a).
All atoms in the graphene sheets were kept at ideal positions. The
in-plane lattice constant $a_0$ is 2.46 \AA, and the inter-plane distance $d$ 
is 3.35 \AA \cite{MSBM}. 
A mixed basis set with 
23 BSs over a range of 5.0 $a_0$ 
and 2D PW with an energy cutoff of 30 Ry 
was used to expand the wavefunction.   

In the absence of the electric field, bilayer graphene 
is a gapless semiconductor. 
Previous works \cite{Mc,MF} with a tight-binding model have reported that 
the application of an external electric field to the bilayer system
with AB-stacking lifts the degenerate states at K to 
form the so-called Mexican-hat-like band structure. The 
field-induced energy gap occurs not at K but slightly away from it. This 
means that a more dense $k$-point sampling near K valley is required to 
ensure the calculation convergence. 
Here, instead of a  
uniform division, we use a non-uniform
$k$-point distribution, as demonstrated in Fig. \ref{ab_E}(b). 
In essence, there are three zones of different sampling density in the BZ,
with denser sampling near K point.  
For clarity, we denoted the division in 
Fig. \ref{ab_E}(b)  as $3 \times 3 \times 3$. 
We found that the sampling with
$12 \times 6 \times 6$ division leads to excellent convergence. 
This corresponds to only 31 $k$-points in IBZ, as compared to 184 points 
by the uniform $44 \times 44$ Monkhorst-Pack
division used in literature \cite{SK}.

\begin{figure}[h]
        \caption{ } \label{hat}
\includegraphics{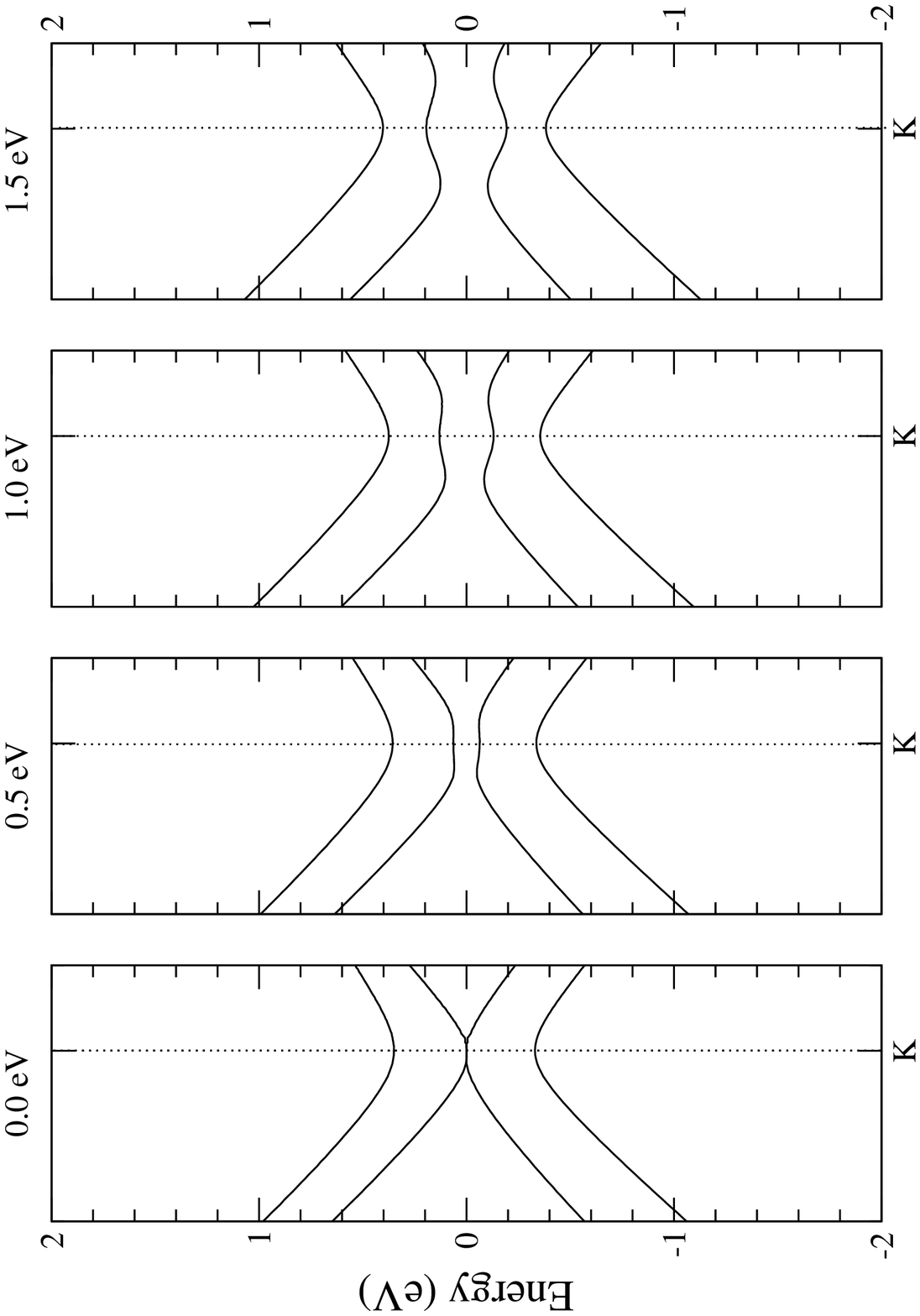}
\end{figure}

To compare with the earlier DFT calculation \cite{MSBM}, we change the strength
${\mathcal{E}}_0$ such that the parameter $U_{ext}= {\mathcal{E}}_0 d$ 
is 0, 0.5, 1.0, and 1.5 eV. The band structures of these four cases near 
K valley (along fragment of $\Gamma$-K-M ) are shown in Fig. \ref{hat}. 
Clearly, the gap increases with increasing $U_{ext}$. 
It is found that a nice agreement with previous 
results \cite{MSBM} was obtained.  This agreement confirms that our 
MBA successfully captures the above characters 
of the complicated band structure near Fermi level at K valley.
\subsection{Dielectric constant of MS$_2$ (M=Mo, Zr) under an external E-field}
Finally, we studied the out-of plane static dielectric constant of 2D TMDC  
under an external electric field. 
The electronic property of MoS$_2$ is influenced by atomic 
structure and can be controlled by various ways 
\cite{CLTRB}-\cite{LLGCL}. 
Layered MoS$_2$ has been considered a promising
candidate for the post-silicon-era field-effect-transistor \cite{VDR}. 
We select MoS$_2$ and ZrS$_2$ to study  
the field-induced dipole moment of their monolayer (1L) and  bilayer (2L)  
to simulate the influence of the gate voltage upon such nanoelectronic devices. 

The $z$-coordinate of all ions was allowed to relax.
The Van der Waals interlayer interactions were taken into account 
with Grimme's 
DFT-D2 version \cite{DFT-D2} to correct the total energy and force. 
The energy cutoff of 2D PW was increased to 40 Ry.
There are 45 BSs over a range of 6$a_0$ for the 2L case. 
In addition, the amplitude ${\mathcal{E}}_0$ 
is chosen such that 
the electrostatic potential $V_E$ at the edge region in the $z$ 
direction   
is higher than the Fermi level to avoid  
unwanted charge transfer from the surface to the edge. 
To compare with the literature \cite{LPV}, 2H MoS$_2$ stacking in A-B order 
and 1T ZrS$_2$ in A-A order for the 2L case are considered. 
Here, for the 1T phase, the two S atoms in one unit cell occupy at the sites 
$(1/3, 2/3, u)$ and $(2/3, 1/3,-u)$. 
The surface $k$-point sampling is the same with the first example. 

The polarized layered structure exhibits
a net electric dipole moment $m$ in response to the external E-field,   
\begin{equation}
m = \int dz \ \bar{n}_{scr}(z)\ z,
\end{equation}
where $\bar{n}_{scr}$ is the screening 
planar-averaged total charge density (including the 
electronic and ion contributions),
\begin{equation}
\bar{n}_{scr}(z) = \frac{1}{A}\int d { \boldsymbol \rho} \
(n_E({\bf r})- n_0({\bf r})), 
\end{equation}
with $n_E$ and $n_0$ being respectively the charge density with and without the 
external field.
We checked the  calculations for 
the 1L MoS$_2$ case with various ${\mathcal{E}}_0$  
and found that $m$ changes linearly in ${\mathcal{E}}_0$, as expected. 

\begin{figure}[h]
        \caption{ } \label{pol}
\includegraphics{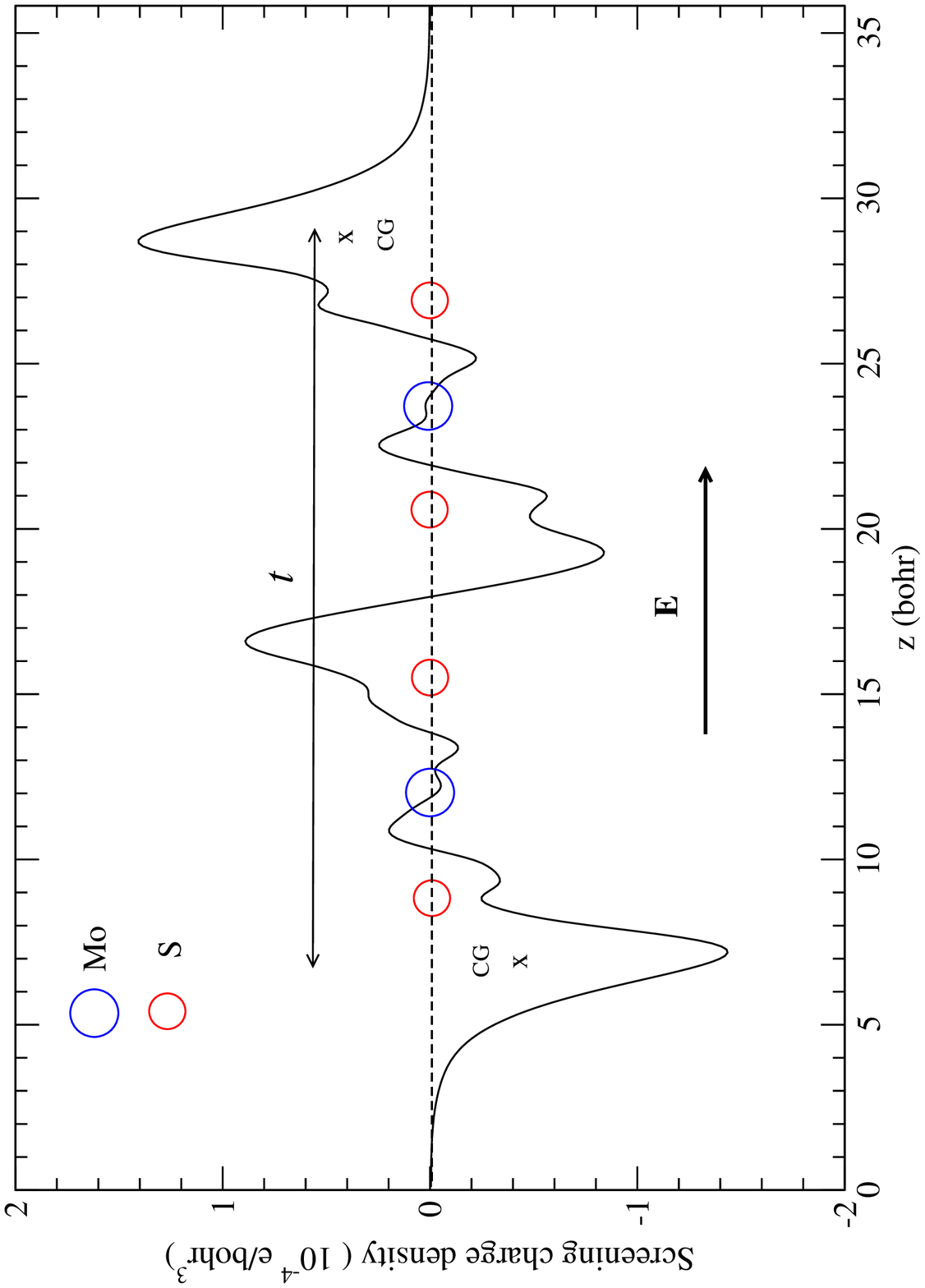}
\end{figure}

The static dielectric constant $\epsilon_{0}$ was calculated via  
\begin{equation}
\epsilon_{0} = \frac{{\mathcal{E}}_0}{{\mathcal{E}}_0-4\pi {\text P}}. 
\label{die}
\end{equation}
The average polarization ${\text P}$ here was deduced from $m/t$ \cite{MV}
where $t$ is the effective thickness of the layers interested. Obviously,  
the dielectric constant is sensitive to the thickness 
chosen. For example, when we use the thickness suggested in 
Ref. \cite{LPV}, the dielectric constant of 1L and 2L MoS$_2$ 
is respectively, 6.2 and 6.8. But, if the thickness were chosen as
the distance between the center of gravity of the screening  
charge distribution of the top and bottom charge layers suggested in 
Ref. \cite{MV} (also shown in Fig. \ref{pol}), then these values become 
38.9 and 10.4, respectively. 
 
Table \ref{tab2} summarizes the calculated dielectric constant 
 and the thickness used.  
These results were evaluated based upon ${\mathcal{E}}_0=0.005$ a.u.. 
Since the dielectric constant can be rescaled for
different values of thickness, we also present the  
value of $m$ as the base for calculations. 
It is clear that the dielectric constant obtained with the thickness from 
Ref. \cite{LPV} is in agreement with that work \cite{LPV}. So,
our approach can directly calculate the dielectric through 
Eq. (\ref{die}) without the correction due to the existence of vacuum spaces
needed in SCA \cite{SK,LPV}.  

To avoid ambiguity in the dielectric constant with  
the choice of thickness, we address this issue from another approach.   
In analog with the atomic Stark effect, the external E-field can significantly 
modulate the band gap of bilayer graphene
 and TMDC \cite{RNT}-\cite{ISACZ}. 
This indicates that the change of the band gap could reflect information of
the screened electric field inside the bilayer system, which is the central 
point in calculating the dielectric constant. Based on this observation, 
we suppose that 
the reduction in the gap from the ${\mathcal{E}}_0=0$ case is approximately 
linear to the amount of the net electric field strength. In this respect,  
we additionally performed a non-self-consistent calculation of 
the gap difference $\Delta E_{g,nscf}$ under the external E-field 
by keeping the charge density $n_0$ unchanged as at zero field. 
Together with the gap difference $\Delta E_{g,scf}$ obtained by using 
the self-consistent charge density $n_E$ in response to the E-field, we can 
estimate the dielectric constant as the ratio of 
$\Delta E_{g,nscf}$ to $\Delta E_{g,scf}$. Through this scenario, the 
dielectric constant of 2L MoS$_2$ and ZrS$_2$ is found to be 9.0 and 4.5,
respectively, comparable to the values of 10.4 and 5.1 
in Table II obtained with the 
thickness suggested in Ref. \cite{MV}. Therefore, the use of such a ratio could
reasonably determine the dielectric constant and it 
does not depend on the effective thickness $t$.  

\section{CONCLUSIONS} \label{con}
In conclusion, we have successfully implemented MBA 
to investigate the atomic and electronic structures of 2D systems
by expanding the wavefunctions with PW for the periodic directions and BS for
the non-periodic direction. Contrary to the existing algorithms based
upon SCA with repeated slabs embedded in vacuum regions, 
MBA is a real space approach along
the non-periodic direction. For charged systems due to carrier injection, 
the spurious Coulomb interaction between the carrier, its images and 
the compensating background charge by SCA is 
avoided in MBA. Moreover, we can directly study the charge polarization of 
the system influenced under an external E-field with no
potential discontinuity that appears in SCA. 
The localized BS has proved to be flexible for expanding the out-of-plane 
wavefunction during geometry optimizations. 
Our MBA results for the atomic relaxation and dielectric constant under 
the applied E-field are consistent with SCA calculations when 
both approaches are equally applicable. However, it is found that 
SCA is no longer valid for treating the charged system 
with high carrier density 
by imposing artificial charge neutrality condition.
We believe that MBA-BS  
is suitable for investigating more realistic 2D materials. 
\begin{acknowledgments}
This work was supported by Ministry of Science and Technology under 
grant numbers MOST 108-2112-M-017 -001 and MOST 108-2112-M-001-041 
and by National Center for Theoretical Sciences of Taiwan.
\end{acknowledgments}

\section{Appendix I}
\subsection{construction of $H|\phi>$}
In MBA, we recall that the wavefunction is expanded as
\begin{equation}
| \phi_{\bf k}>
= \sum_{j{\bf g}} C_{j,{\bf k+g}} |{ \bf k +  g} ; j, \kappa>.
\end{equation}
We will omit index ${\bf k} $ in $| \phi_{\bf k}> $ hereafter.
\subsubsection{kinetic energy part}
This part is given by
\begin{align}
  - \nabla^2\  | \phi>
\nonumber\\
=& \sum_{\bf g} \sum_{|j-j'|\le \kappa} C_{j,{\bf k+g}} 
(<  {B'}_{j',\kappa} | \ {B'}_{j,\kappa}>\
+
<  B_{j',\kappa} |\ B_{j,\kappa}> )
 |{\bf k + g}|^2  
|{ \bf k +  g} ; j', \kappa>.	\label{eqkin}
\end{align}
$ {B'}_{j,\kappa}$ in Eq. (\ref{eqkin}) means the derivative of 
$ B_{j,\kappa}$ with respective to the $z$ coordinate. 
\subsubsection{local potential part}
In practical calculations, we separate the local potential of atomic PP into a  
long-range potential, $V^{at,l}_{loc}(r)=-\frac{Z}{r}{\rm erf}(\sqrt{a}r)$ 
and a short-range remainder,   
\begin{equation}
V^{at}_{loc}(r)= V^{at,l}_{loc}(r) +V^{at,s}_{loc}(r).
\end{equation}
The former corresponds to the potential due to an 
auxiliary charge distribution 
\begin{equation}
n_a({\bf r})=(\frac{a}{\pi})^{3/2}Z e^{-a |{\bf r}|^2}	\label{eqauxa}.
\end{equation}
Using the fact that the 2D Fourier transform of the Coulomb potential is
\begin{equation}
\int d z' n_a({\bf g},z')
\frac{2 \pi}{|{\bf g|}} \
e^{- |{\bf g}| |z-z'|}, 
\end{equation}
it is straightforward that the long-range part of $V_{loc}$ in Eq. (\ref{veff})
in 2D momentum representation can be written as  
\begin{eqnarray}
V^l_{loc}({\bf g} \neq 0,z)&& = 
\frac{1}{A}\sum_I \frac{ \pi Z_I }{|{\bf g}|} \left ( 
                  e^{ |{\bf g}| |z-z_I|} {\rm erfc}
			(\sqrt{a}( |z-z_I|+\frac{|{\bf g}|}{2a})) 
		  +e^{-|{\bf g}| |z-z_I|} {\rm erfc}
			(\sqrt{a}(-|z-z_I|+\frac{|{\bf g}|}{2a})) 
			\right )  \nonumber \\
		&&	\times e^ {-i{\bf g} \cdot { \bf \rho_I}}, \label{eqlocgn0} \\
V^l_{loc}({\bf g} = 0,z) && = 
-\frac{1}{A}\sum_I  2\pi Z_I 
\left (|z-z_I| {\rm erf} (\sqrt{a}|z-z_I|)+ \frac{1}{\sqrt{a\pi}}e^{-a|z-z_I|^2} 
\right ).	\label{eqlocg0} 	
\end{eqnarray}
%

For the short-range part $V^s_{loc}( {\bf g},z)$,  
we first calculate 
\begin{equation}
V^{at,s}_{loc}( {\bf G} )= 		
\int d{\bf r} e^{-i {\bf G} \cdot { \bf r}} V^{at,s}_{loc}( { \bf r}) \label{eqlocs0}\\
\end{equation}
and 
\begin{equation}
V^{s}_{loc}( {\bf G} )=
 \sum_{I}e^{-i {\bf G \cdot R_I}} V^{at,s}_{loc}( {\bf G} ).    \label{eqlocs1} 
\end{equation}
${ \bf G }$ is a compact notation for $({\bf g},g_z)$ with $g_z=2\pi n/L $.
Here  $n$ is a integer and $L$ is the height along $z$ direction. 
It follows that 
$V^s_{loc}( {\bf g},z$) can be obtained by fast Fourier transform (FFT) from
\begin{equation}
V^s_{loc}( {\bf G} )= \frac{1}{L} \int dz e^{-i g_z z}
V^s_{loc}( {\bf g},z).     \label{eqlocs}                  
\end{equation}
Together with the exchgane-correlation potential
\begin{equation}
V_{xc}( {\bf g},z)= \frac{1}{A}\int d {\bf \rho} 
e^{ -i  \bf  g \cdot {\rho}}V_{xc}({ \bf r}),  	
\end{equation}
we have $V_{\text{eff}}$ in ${\bf g}$-space representation.

\subsubsection{nonlocal potential part}
$Q^I_{nm}({\bf g},z)$ was calculated by analogy with the $V^{s}_{loc}$ case. 
By knowing both 
$Q^I_{nm}({\bf g},z)$ and $V_{\text{eff}}({\bf g},z)$, we can update
$D^I_{nm}$ in Eq. (\ref{eq1}) in each interation. Furthermore,  
\begin{eqnarray}
&& < { \bf k +  g } ; j, \kappa | \beta^I>   	\nonumber	\\
&=&
e^{ -i \bf (k +  g) \cdot {\rho_I} }
\int d{\bf r}
 B_{j,\kappa}(z+z_I) e^{ -i  \bf (k +  g) \cdot {\rho}} \beta(r)
Y_{lm}({\bf\hat{r}})\nonumber \\	
&=& \frac{2\pi}{i^m} \sqrt{\frac{(2l+1)(l-m)!}{4\pi(l+m)!}}
e^{ -i  \bf (k +  g) \cdot {\rho_I}}
\int r^2 dr \sin\theta d\theta
\ B_{j,\kappa}(r\cos\theta+z_I) J_m(|{ \bf \ k + g }|r\sin\theta)
P_l^m(\cos\theta)\beta(r)	\nonumber	\\
	\label{eqb}	
\end{eqnarray}
where
$Y_{lm}$ is the spherical harmonics,
$P_l^m$ is the associated Legendre function, and $J_m$ is Bessel
function of order $m$. Using this one can evaluate $ < \beta^I_m|\phi> $ and therefore
$V_{NL}|\phi>$ 
\begin{equation}
V_{NL} | \phi> \ = \sum_{Inm}D^I_{nm} < \beta^I_m|\phi> |\beta^I_n>. 
\end{equation}

With $H|\phi>$ and $S|\phi>$ in hand, 
we perform the eigenvalue/eigenvector searching. 
To speed up calculations, we expand  
$H|\phi>$/$S|\phi>$ as well as $|\phi>$ in terms of a complementary   
basis $ | { \bf k +  g} ; j, \kappa >_o $ where the B-splines  
$B^o_{j,\kappa}$ are orthogonalized (e.g. via the Grand-Schmidt procedure).
Particularly, 
the expansion for $V_{\text{eff}}|\phi>$ was obtained by 2D FFT 
with the computation of 
$\int dz V_{\text{eff}}({\boldsymbol {\rho}},z) \phi({\boldsymbol {\rho}},z) 
B^o_{j,\kappa}(z)$. 
Therefore,
the vector-product performance in the conjugate-gradient algorithm 
is similar to that implemented in the planewave-based codes  \cite{KF,bfgs}.

Of course, $V_E(z)$ in Eq. (\ref{eq23})
is added to $H$ if an external E-field is applied.  
In practice, the reference electric potential is set at 
the middle height in the $z$ direction.
\subsection{force:}
\subsubsection{component due to the local potential}
From Eqs. (\ref{eqlocgn0}) and (\ref{eqlocg0}), the 
derivative of the long-range part can be shown as 
\begin{eqnarray}
&& \frac{\partial V^l_{loc}({\bf g} \neq 0,z)}{\partial z_I}
   \nonumber \\
&=& \left \{ \frac{ \pi Z_I}{A} \left (
                  e^{ |{\bf g}| (z-z_I)} {\rm erfc}
                        (\sqrt{a}( (z-z_I)+\frac{|{\bf g}|}{2a}))
                 -e^{-|{\bf g}| (z-z_I)} {\rm erfc}
                        (\sqrt{a}(-(z-z_I)+\frac{|{\bf g}|}{2a}))
                        \right )  \right.  \nonumber \\
&+&  \frac{  \pi Z_I }{A|{\bf g}|} \sqrt{a/\pi} \left . \left (
                  e^{ |{\bf g}| (z-z_I)} e^
                        {-a( (z-z_I)+\frac{|{\bf g}|}{2a})^2}
                  - e^{-|{\bf g}| (z-z_I)} e^
                        {-a(-(z-z_I)+\frac{|{\bf g}|}{2a})^2}
                       \right )  \right \} e^{-{\bf g} \cdot {\bf \rho_I}}, \\
&& \frac{\partial V^l_{loc}({\bf g} = 0,z)}{\partial z_I} = 
-\frac{2 \pi Z_I}{A} {\rm erf}(\sqrt{a}(z-z_I)).
\end{eqnarray}
Equation (\ref{eqlocs1}) and (\ref{eqlocs}) imply 
\begin{equation}
-ig_z V^{at,s}_{loc}( {\bf G} )= \frac{1}{L} \int dz e^{-i g_z z}
\frac{ d V^s_{loc}( {\bf g},z)}{d z_I},	\label{eqlocsf}                        
\end{equation} 
and from this we can obtain the short-range part 
$ \partial V^s_{loc}( {\bf g},z)/\partial z_I$ by FFT. 
Then the force component in Eq. (\ref{eq10}) was evaluated by 
\begin{equation}
F^{loc}_{Iz}= -A_c\sum_{{\bf g}} \int dz n^*( {\bf g},z)(
\frac{\partial V^l_{loc}({\bf g},z)}{\partial z_I}+
\frac{\partial V^s_{loc}({\bf g},z)}{\partial z_I}),
\end{equation}
where $A_c$ is the unit-cell area.
\subsubsection{component due to the nonlocal potential}
In the momentum space, $Q_{nm}^{at}$ was expressed as 
\begin{equation}
Q_{nm}^{at}( {\bf G} )=
\int d{\bf r} e^{-i {\bf G} \cdot { \bf r}} Q_{nm}^{at}( { \bf r}) =
\sum_{LM}
C_{LM}^{l_nm_n,l_mm_m} 
\int d{\bf r} e^{-i {\bf G} \cdot { \bf r}} Q_{nm}^{at,L}(r)
Y_{LM}({\bf\hat{r}}).	
\end{equation}
Here, $C_{LM}^{l_nm_n,l_mm_m}$ 
is Clebsch-Gordan coefficient. In this work, we only consider the lowest
angular moment \cite{vancode}. 
Following the similar procedures given in Eqs. (\ref{eqlocs0})-(\ref{eqlocs}) 
and (\ref{eqlocsf}), we get $\partial Q_{nm}^{I}( {\bf g},z)/\partial z_I$.  
With these results,  
the integral in Eq. (\ref{eq11}) was performed in ${\bf g}$-space to 
yield $F^{nl,1}_{Iz}$.

Next, it is clear that 
\begin{eqnarray}
&& < { \bf  k +  g } ; j, \kappa |\frac{\partial \beta^I}{\partial z_I} > 
\nonumber  \\
&=& \frac{2\pi}{i^m} \sqrt{\frac{(2l+1)(l-m)!}{4\pi(l+m)!}}
e^{ -i  \bf (k +  g) \cdot {\rho_I}}
\int r^2  dr \sin\theta d\theta
\ B^{'}_{j,\kappa}(r\cos\theta+z_I) J_m(|{ \bf \ k + g }|r\sin\theta)
P_l^m(\cos\theta)\beta(r)       \nonumber       \\
\end{eqnarray}
through Eq. (\ref{eqb}). Consequently, 
we have 
$<\phi| \frac{ \partial \beta^I}{\partial z_I}>$
and $F^{nl,2}_{Iz}$ in Eq. (\ref{eq12}) can be evaluated.  
\subsubsection{component due to NLCC}
In a similar way for  
$ \partial V^s_{loc}( {\bf g},z)/\partial z_I$, we calculate 
$ \partial n_{c}( {\bf g},z)/\partial z_I$ to obtain 
\begin{equation}
F^{nlcc}_{Iz}=-A_c 			
\sum_{\bf g} \int d z v^*_{xc}({\bf g},z) 
\frac{\partial n_c^{I}( {\bf g},z)}{\partial z_I} . 
\end{equation}
\subsubsection{component due to the ion-ion interaction}
For the 2D case, the ${\bf g}$-space part of the 
Ewald sum for $E_{ii}$ with the auxiliary charge distribution 
\begin{equation}
n_{\bar{a}}({\bf r})=(\frac{\bar{a}}{\pi})^{3/2}Z e^{-\bar{a}|{\bf r}|^2}	\label{eqauxa1}
\end{equation}
is \cite{LCnote}
\begin{eqnarray}
&& E^k_{ii} 	\nonumber \\
&=& \frac{1}{A}\sum_{I \neq J} \left \{ \sum_{{\bf g \neq 0}} \frac{ \pi Z_I Z_J }{|{\bf g}|} \left (
                  e^{ |{\bf g}| |z_I-z_J|} {\rm erfc}
                        (\sqrt{\bar{a}}( |z_I-z_J|+\frac{|{\bf g}|}{2\bar{a}}))
                  +e^{-|{\bf g}| |z_I-z_J|} {\rm erfc}
                        (\sqrt{\bar{a}}(-|z_I-z_J|+\frac{|{\bf g}|}{2\bar{a}}))
                        \right ) \right.  \nonumber \\
&& \left.  + Z_I Z_J 
\left (\pi |z_I-z_J| {\rm erf} (\sqrt{\bar{a}}|z_I-z_J|)+ 
\sqrt{\pi/\bar{a}}e^{-\bar{a}|z_I-z_J|^2} \right ) \right \} 
                \  \cos({\bf g} \cdot ({\bf \rho_I-\rho_J})). 
\end{eqnarray}
After some algebra, the corresponding force is found to be 
\begin{eqnarray}
&& F^{ii}_{k,Iz}  \nonumber \\
&=&\frac{Z_I}{A}\sum_ J \left \{  \sum_{{\bf g} \neq 0} 
 \left [ \pi Z_J \left (
                  e^{ |{\bf g}| (z_I-z_J)} {\rm erfc}
                        (\sqrt{\bar{a}}( (z_I-z_J)+\frac{|{\bf g}|}{2\bar{a}}))
                 -e^{-|{\bf g}| (z_I-z_J)} {\rm erfc}
                        (\sqrt{\bar{a}}(-(z_I-z_J)+\frac{|{\bf g}|}{2\bar{a}}))
                        \right )  \right. \right. \nonumber \\
&+&  \frac{ 2 \pi Z_J }{|{\bf g}|} \sqrt{\bar{a}/\pi} \left. \left . \left (
                  e^{ |{\bf g}| (z_I-z_J)} e^
                        {-a( (z_I-z_J)+\frac{|{\bf g}|}{2\bar{a}})^2}
                  - e^{-|{\bf g}| (z_I-z_J)} e^
                        {-\bar{a}(-(z_I-z_J)+\frac{|{\bf g}|}{2\bar{a}})^2}
                       \right )  \right ] 
		-2 \pi Z_I {\rm erf}(\sqrt{\bar{a}}(z_I-z_J)) \right \}\nonumber \\
                && \times    \cos({\bf g} \cdot ({\bf \rho_I-\rho_J})). 
\end{eqnarray}
For the sake of completeness, we write down the ${\bf r}$-space part 
of Ewald sum and its force component 
$F^{ii}_{r,Iz}$
\begin{equation}
E^r_{ii} = \frac{1}{2} \sum_{I \neq J}  Z_I Z_J 
\frac{{\rm erfc}(|{\bf R_I}-{\bf R_J}|\sqrt{\bar{a}})}{|{\bf R_I}-{\bf R_J}|}, 
\end{equation}
\begin{equation}
F^{ii}_{r,Iz} = -Z_I\sum_{J} Z_J  \left (
2\sqrt{\bar{a}/\pi}e^{-\bar{a}{|\bf R_I}-{\bf R_J}|^2}+\frac
{{\rm erfc}(|{\bf R_I}-{\bf R_J}|\sqrt{\bar{a}})} 
{|{\bf R_I}-{\bf R_J}|} \right )
\frac{ z_I-z_J}{|{\bf R_I}-{\bf R_J}|^2}. 
\end{equation}
In this work, we set $\bar{a}$ in Eq. (\ref{eqauxa1}) equal to 
$a$ in Eq. (\ref{eqauxa}).
%
\subsubsection{components due to the external E-field/Van der Waals interaction}

When the system is influenced under an applied E-field, 
the associated force component is 
\begin{equation}
F^E_{Iz}=-\frac{
\partial E_E}{\partial z_I } =Z_I{\mathcal{E}}_0.
\end{equation}
Moreover, if the Van der Waals interaction were taken into account, 
the DFT-D2 dispersion 
pair energy \cite{DFT-D2} of
\begin{equation}
E_{disp} =-\frac{1}{2} s_6\sum_{I \neq J}   
\frac{C_{6,IJ}}{|{\bf R_I}-{\bf R_J}|^6}\ f(|{\bf R_I}-{\bf R_J}|) 
\end{equation}
with a damping factor 
\begin{equation}
f_{IJ}=f(|{\bf R_I}-{\bf R_J}|)=\frac{1}{1+e^{-d(|{\bf R_I}-{\bf R_J}|/R_{IJ}-1)}},   
\end{equation}
is further added to the total energy. Note that the Ewald sum technique 
\cite{KSS} is not implemented here.  
The parameters $s_6$,$C_{6,IJ}$, $R_{IJ}$ and $d$ can be found in Ref. \cite{DFT-D2}.
It can be easily shown that the 
$z$-component force due to this dispersion correction is given by 
\begin{equation}
F^{disp}_{Iz} =- s_6\sum_{J}   
C_{6,IJ}f_{IJ} 
\left [ -f_{IJ}e^{-d(|{\bf R_I}-{\bf R_J}|/R_{IJ})} 
\frac{d}{R_{IJ}}+\frac{6}{|{\bf R_I}-{\bf R_J}|}\right ] \frac{z_I-z_J}
{|{\bf R_I}-{\bf R_J}|^7}.
\end{equation}
The $x-$ or $y-$ components require no further comment. 
The details for these two components are described elsewhere \cite{Kax,Koh}.
\newpage
\begin{center}
\large {\bf FIGURE CAPTIONS}    \normalsize
\end{center}
Fig. 1: Total energy versus internal coordinate $u$ of S in MoS$_2$ monolayer. 
	The arrow indicates the energy minimum predicted by 
	Broyden-Fletcher-Goldfarb-Shanno (BFGS) algorithm with $u=0.4972$.  
	See text for details. \\ \\
Fig. 2: Band structure of (a) MoS$2$ monolayer with a direct gap and 
	(b) ZrS$2$ monolayer with an indirect gap. (c) and (d) are the 
        corresponding VASP counterparts.	\\ \\ 
Fig. 3: Band structure of charged graphene (a) without and (b) with 
        charge neutrality condition by the present method. 
        (c) Counterpart by VASP with charge neutrality condition.  
	\\ \\ 
Fig. 4: Convergence of the effective local potential during iteration 
	for both electrically neutral and charged graphene.  \\ \\
Fig. 5: (a) Graphene bilayer with A-B stacking under an external E-field. 
	(b) Non-uniform $k$-point sampling in surface Brillouin zone  	
	for the calculations in (a). \\ \\
Fig. 6: Band structure of graphene bilayer near K 
        under an external electric potential with various strengths. 
	See text for details. \\ \\
Fig. 7: (Color online) Screening charge density distribution of MoS$2$ bilayer  
        infleunced by an external E-field.
	Thickness $t$ defined as the distance between 
        the center of gravity (CG) of the screening  
	charge distribution of outermost charge layers \cite{MV}. 
\newpage
\newpage
\newpage
\newpage
\newpage
\newpage
\newpage
\newpage
\newpage
\begin{table}[p]
        \caption{ Relevant information for the total number of the
		  wavefunction basis at point K used in 
		  MBA and VASP for monolayer MoS$_2$ and
	 	  graphene (G). The data are produced based on lattice 
		  constant $a_{0}$ of 3.16 and 2.46 \AA~ and
		  energy cutoff of plane waves of
		  20 and 30 Ry for MoS$_2$ and graphene, respectively. 
		  The value labeled '$z$-range' (\AA) means the range 
		  where B-splines $B_{j,\kappa}$ are
		  distributed. Two sets of $c$ values (lattice constant along 
		  the perpendicular $z$ direction, \AA) with vacuum layer of
		  10 and 15 \AA~
		  required in VASP are listed for comparison. } \label{tab0}
        \normalsize
        \begin{center}
                \begin{tabular}{llllllll}
\hline  \hline
  \hspace{15mm} &MBA & & &VASP & & & 	\\
\hline  
  \hspace{15mm} &$z$-range    \hspace{10mm} 
                &$B_{j,\kappa}$ \#   \hspace{10mm} 
                &basis \#   \hspace{10mm} 
                &$c$        \hspace{10mm} 
                &basis \#   \hspace{10mm} 
                &$c$        \hspace{10mm} 
                &basis \#   \hspace{10mm} \\ 	
\hline  
MoS$_2$  &12.64 (4$a_0$)   &25   &1100   &13.16    &1218   &18.16   &1662  \\   
G        &8.00 (3.25$a_0$) &13   &546    &10.00    &966   &15.00   &1422  \\   
\hline \hline
                \end{tabular}
        \end{center}
\end{table}
\newpage
\begin{table}[p]
        \caption{ The optimized vertical M-X distance $d_z$(\AA) and band gap 
                  $E_g$ (eV) in 2H MX$_2$ (M=Mo, W, Zr, Hf; X=S, Se) 
		  monolayer with experimental lattice 
                  constant $a_0$ (\AA).  VASP results are also shown for 
                  comparison.}  \label{tab1}
        \normalsize
        \begin{center}
                \begin{tabular}{lllllllll}
\hline  \hline
\hspace{8mm} &MoS$_2$   
\hspace{8mm} &WS$_2$  
\hspace{8mm} &MoSe$_2$  
\hspace{8mm}   &WSe$_2$        
\hspace{8mm}   &ZrS$_2$        
\hspace{8mm}   &HfS$_2$        
\hspace{8mm}   &ZrSe$_2$        
\hspace{8mm}   &HfSe$_2$        \\
\hline 
$d_z$     &1.57              &1.58            &1.68    &1.68 &1.56 &1.53 &1.68&1.65       \\
$d_z$(VASP) &1.57            &1.59            &1.68    &1.70 &1.56 &1.54 &1.68&1.66      \\
\hline 
$E_g$     &1.79              &1.97            &1.58    &1.68 &0.95 &0.96 &0.81
&0.82      \\
$E_g$(VASP) &1.80            &1.97            &1.56    &1.68 &0.95 &1.04 &0.79
&0.88      \\
\hline 
$a_0$(exp)
&3.16\footnote{ Ref. \cite{SHM}}          
&3.15\footnote{ Ref. \cite{SBJ}}       
&3.28\footnote{ Ref. \cite{APJ}}   
&3.28\footnote{ Ref. \cite{SBJ}}  
&3.66\footnote{ Ref. \cite{SBJ}}        
&3.63\footnote{ Ref. \cite{HS}}        
&3.75\footnote{ Ref. \cite{GR}}   
&3.77\footnote{ Ref. \cite{HS}}  \\
\hline \hline
                \end{tabular}
        \end{center}
\end{table}
\newpage
\begin{table}[p]
        \caption{ The static out-of-plane dielectric constant $\epsilon_{0}$ of
		  monolayer (1L) and bilayer (2L) MoS$_2$ and ZrS$_2$  
		  using various effective 
		  thickness $t$ (a.u.). The dipole moment $m$ multiplied by 
		  $4\pi$ ($\times 10^{-2}$ e/bohr$^2$) is also listed. 
		  See text for details.} \label{tab2}
        \normalsize
        \begin{center}
                \begin{tabular}{lllllll}
\hline  \hline
  \hspace{15mm} &$4\pi m$  \hspace{15mm} &$t$\footnote{Ref. \cite{LPV}} 
                \hspace{15mm} &$\epsilon_0$ 
		\hspace{15mm} &$t$\footnote{calculated by following 
		Ref. \cite{MV}}
                \hspace{15mm} &$\epsilon_0$ 	
		\hspace{15mm} &${\epsilon_0}$\footnote{Ref. \cite{LPV}} 	
\\ 	
\hline  
MoS$2$  &   &   &        &   &   &     \\   
1L      &4.86   &11.57   &6.3        &9.97   &38.9   &6.4     \\   
2L        &9.88   &23.12   &6.9        &21.86   &10.4   &6.8     \\   
\hline  
ZrS$2$  &   &   &        &   &   &     \\   
1L        &4.65   &10.85   &7.0        &10.51   &8.7   &6.8     \\   
2L        &9.31   &21.66   &7.1        &23.12   &5.1   &6.9     \\   
\hline \hline
                \end{tabular}
        \end{center}
\end{table}

\begin{thebibliography}{99}
\bibitem{note1}
In our previous work \cite{RCH}, the comparison between MBA and SCA for 1D 
charged system was made only with the charge neutrality condition.
\end{thebibliography}
\end{document}